\documentclass[12pt,onecolumn,preprint]{aastex}

\doublespace
\onecolumn

\newcommand{\etal}{{et~al.~}}
\newcommand{\fcrao}{\mbox{$\rm FCRAO$}}
\newcommand{\IRAS}{\mbox{$\rm IRAS$}}
\newcommand{\HIRES}{\mbox{$\rm HiRes$}}

\newcommand{\co}{\mbox{$^{12}$CO}}

\newcommand{\kms}{\mbox{${\rm km~s}^{-1}$}}
\newcommand{\htwo}{\mbox{${\rm H}_2$}}
\newcommand{\msun}{\mbox{$M_\odot$~}}

\received{}

\lefthead{Heyer \etal}
\righthead{Molecular Gas and the Schmidt Law in M33}

\begin{document}

\def\gtabouteq{\,\hbox{\raise 0.5 ex \hbox{$>$}\kern-.77em
                    \lower 0.5 ex \hbox{$\sim$}$\,$}}
\def\ltabouteq{\,\hbox{\raise 0.5 ex \hbox{$<$}\kern-.77em
                     \lower 0.5 ex \hbox{$\sim$}$\,$}}
\def\vlsr{V$_{LSR}$}
\def\kms{km s$^{-1}$}

\title{The Molecular Gas Distribution and 
Schmidt Law in M33}

\author{Mark H. Heyer\altaffilmark{1}, 
Edvige Corbelli\altaffilmark{2},
Stephen E. Schneider\altaffilmark{1}, 
Judith S. Young\altaffilmark{1}
}

\altaffiltext{1}{Department of 
Astronomy, Lederle Research Building,
University of Massachusetts, Amherst, MA 01003, USA;
heyer@astro.umass.edu,schneider@astro.umass.edu,young@astro.umass.edu}

\altaffiltext{2}{INAF-Osservatorio Astrofisico di Arcetri, Largo E. Fermi,
5 I-50125, Firenze, Italy; edvige@arcetri.astro.it}

\begin{abstract}
The relationship between the star formation rate and surface density of 
neutral gas within the disk of M33 is examined with new imaging observations  
of \co\ J=1-0 emission gathered with the \fcrao\ 14m telescope and \IRAS\ 
\HIRES\ images of the 60\micron\ and 100\micron\ emission.  
The  Schmidt law, $\Sigma_{SFR} \sim \Sigma_{gas}^n$, is constructed 
using radial profiles of the HI 21cm, CO, and far infrared emission. 
A strong correlation is identified between the star formation rate and 
molecular gas surface density. This suggests that the condensation of 
giant molecular clouds is the limiting step to star formation within the 
M33 disk. The corresponding molecular Schmidt index, $n_{mol}$, is 
1.36$\pm$0.08. The star formation rate has a steep 
dependence on total mass gas surface density, ($\Sigma_{HI}+\Sigma_{H_2}$),
owing to the shallow radial profile of the atomic gas which dominates 
the total gas surface density for most radii.
The disk pressure of the gas is shown to play a prominent role 
in regulating the molecular gas fraction in M33.
\end{abstract}

\parindent=20pt
\keywords{
galaxies: individual (M33)
galaxies: evolution
galaxies: ISM
stars: formation
}
\parindent=20pt

\section{Introduction}

The understanding of galaxy structure and evolution is ultimately linked 
to the understanding of the star formation process.  Star formation 
within galaxies is commonly described by two relationships
which are oversimplifications of complex interstellar processes taking 
place over a large range of spatial scales (Elmegreen 2002).
The first is an empirical relationship 
between the star formation rate per unit area, $\Sigma_{SFR}$,
and gas mass surface density, $\Sigma_{gas}=\Sigma_{HI}+\Sigma_{H_2}$,
$$ \Sigma_{SFR} \propto \Sigma_{gas}^n  \eqno(1) $$
often referred to as the Schmidt law (Schmidt 1959; Kennicutt 1989; 
Wong \& Blitz 2002).  
The value of the Schmidt index, $n$, and its variation with galaxy type and 
environment provide valuable clues to the physical processes that regulate 
star formation in galaxies.
The second relationship simply states that there is a critical gas 
surface density, $\Sigma_c$, below which star formation is suppressed
(Kennicutt 1989; Martin \& Kennicutt 2001). Since stars form within the 
cold, dense, molecular gas phase of the interstellar medium, the critical
surface density may simply reflect the necessary local conditions at which 
molecules are sufficiently self-shielded from dissociating UV radiation.
However, several studies have shown that the critical surface density is 
dependent upon the galaxy rotation curve (Kennicutt 1989; Martin \& Kennicutt
2001) and relate $\Sigma_c$ to the surface density, $\Sigma_{Q}$,
which parameterizes 
the gravitational instability associated with the Coriolis force (Toomre 1964),
$$\Sigma_{Q}={ {c\kappa}\over{\pi G} } \eqno(2) $$
where $c$ is the sound speed and $\kappa$ is the epicyclic frequency. 
Martin \& Kennicutt (2001) demonstrate that $\Sigma_{Q}$ provides an accurate
gauge of the gas column density at which star formation is inhibited
for most of the 32 galaxies in their sample.  

Two galaxies in the sample of Martin \& Kennicutt (2001) do not conform to the 
Toomre condition.  M33 and NGC 2403 are small, low luminosity, low mass 
galaxies.  Both systems 
 are actively forming stars 
throughout their disks yet are sub-critical ($\Sigma_{gas} << \Sigma_{Q}$)
over all radii (Kennicutt 1989; Martin \& Kennicutt 2001). 
For both galaxies, this assessment is based upon limited imaging of the 
molecular gas component within the central region (Wilson \& Scoville 1989; 
Thornley \& Wilson 1995) or the major axis (Young \etal 1995).  
Using wide field imagery of CO emission presented in this study,
Corbelli (2003) has confirmed the 
sub-critical nature of the M33 disk.  Several studies have proposed that 
star formation in galaxies like M33, NGC 2403, and irregular 
galaxies with slowly rising rotation curves, may be regulated by 
gas instabilities modulated by the local shear rate
(Hunter, Elmegreen, \& Baker 1998; Martin\& Kennicutt 2001; Corbelli 2003).  
The critical surface density may also be modulated by the additional 
gravity from either the stellar disk or the dark matter halo 
(Wang \& Silk 1994; Hunter, Elmegreen, \& Baker 1998; Corbelli 2003).

M33 is a nearby galaxy which shows no signs of interaction
with other galaxies.  Therefore, it constitutes an excellent laboratory 
to evaluate the degree to which physical processes regulate star formation 
in disk galaxies.
In this paper we present the first, unbiased, filled aperture census
of the molecular gas component in M33 from an imaging survey of CO
emission with the FCRAO 14m telescope.  A recent high resolution
interferometric CO image of M33 is presented by Engargiola \etal (2003).
The FCRAO data have been previously 
used to derive the radial distributions of molecular
hydrogen, the rotation curve, and the gravitationally unstable region in
the disk of M33 (Corbelli 2003). Here, we examine
the relationship between the star formation rate and gas surface density
as parameterized by equation (1).
It is important to assess whether the star formation properties
(Schmidt Law, star formation efficiency) in sub-critical galaxies
like M33 are modified due to different modes and processes which enable the
condensation of cold, dense clouds from the diffuse ISM.
Moreover, as a low luminosity, molecular poor galaxy,
M33 provides an important complement to the sample of CO-bright spiral 
galaxies used by Wong \& Blitz (2002) to investigate 
the relationship between gas content and star formation.

In $\S$2, the CO data collection, calibration, and processing procedures
are described. The spatial distribution of the molecular gas is presented 
in $\S$3.  In $\S$4, we derive the Schmidt law from HI, CO, and IRAS HiRes
images and evaluate the role of pressure in regulating the molecular 
gas fraction.

\section{FCRAO Observations}

Observations of $^{12}$CO J=1-0 emission from 
M33 were obtained with the 14 meter telescope of the Five
College Radio Astronomy Observatory (\fcrao, full width at half maximum
= 45\arcsec) in March 2000 and between October 2000
and March 2001 using the 16 element focal plane array 
receiver SEQUOIA.  Typical system temperatures were 400 to 700 K.
The backend was comprised of a set of 16 autocorrelation spectrometers
operating at 80 MHz bandwidth with 378 kHz resolution sampled every 313 kHz.
At the observing frequency of the \co\ line (115.2712 GHz), these
spectral parameters correspond to a bandwidth of 
208 \kms\ with a resolution of 0.98 \kms\ sampled at 0.81 \kms. Since this 
bandwidth does not cover the kinematic extent of the 
galaxy, the velocity at which the spectrometer was centered 
varied along the major axis according to 
the HI rotation curve (Corbelli \& Salucci 2000).
For all observed positions within the field, the velocity shift due to 
rotation of M33 within the 6\arcmin\ field of view of the SEQUOIA array
is sufficiently small such that all molecular line emission is fully 
covered within the 80 MHz bandwidth with sufficient baseline window.

To image M33, the focal plane array was aligned along the major axis
with a position angle of 22$^\circ$ (measured East of North) 
about the central J2000 coordinate,
$\alpha,\delta$=01$^h$~33$^m$~50.89$^s$, 30$^\circ$~39$\tt '$~
36.7$\tt ''$.
The data were taken on a 22.1\arcsec\ grid which corresponds to near 
Nyquist sampling.  
Observations were taken in a position switching mode in which 4 consecutive 
source positions were shared with a single off-source position taken at 
the same elevation to ensure the most stable baselines.  
The spectra were calibrated every 20 to 30 minutes with a chopper wheel to 
switch between the sky and an ambient temperature.
All reported intensities refer to main beam brightness 
temperature assuming a main beam efficiency of 0.45.  
In order to obtain more uniform noise across the image and to account for 
infrequent, inoperative pixels within the array, the footprints were 
staggered by half the size of the array along the major axis so that 
any position on the sky was observed by 2 or more pixels.  Due to this 
staggering, the edges of the final image contain less integration time
and therefore, increased noise levels with respect to the median rms value
of the entire field.  The images shown in 
\S3 have been rotated back to the equatorial coordinate system,
resampled to a 25\arcsec\ grid and smoothed with a 22\arcsec\ Gaussian 
kernel to achieve a final angular resolution of 50\arcsec. At this
resolution the median
sensitivity, defined here as the standard deviation of values within 
1 \kms\ wide spectroscopic channels with no emission, is 0.053 K.

\section{Results}
\subsection{Molecular Gas Distribution}

\begin{figure}
\plotone{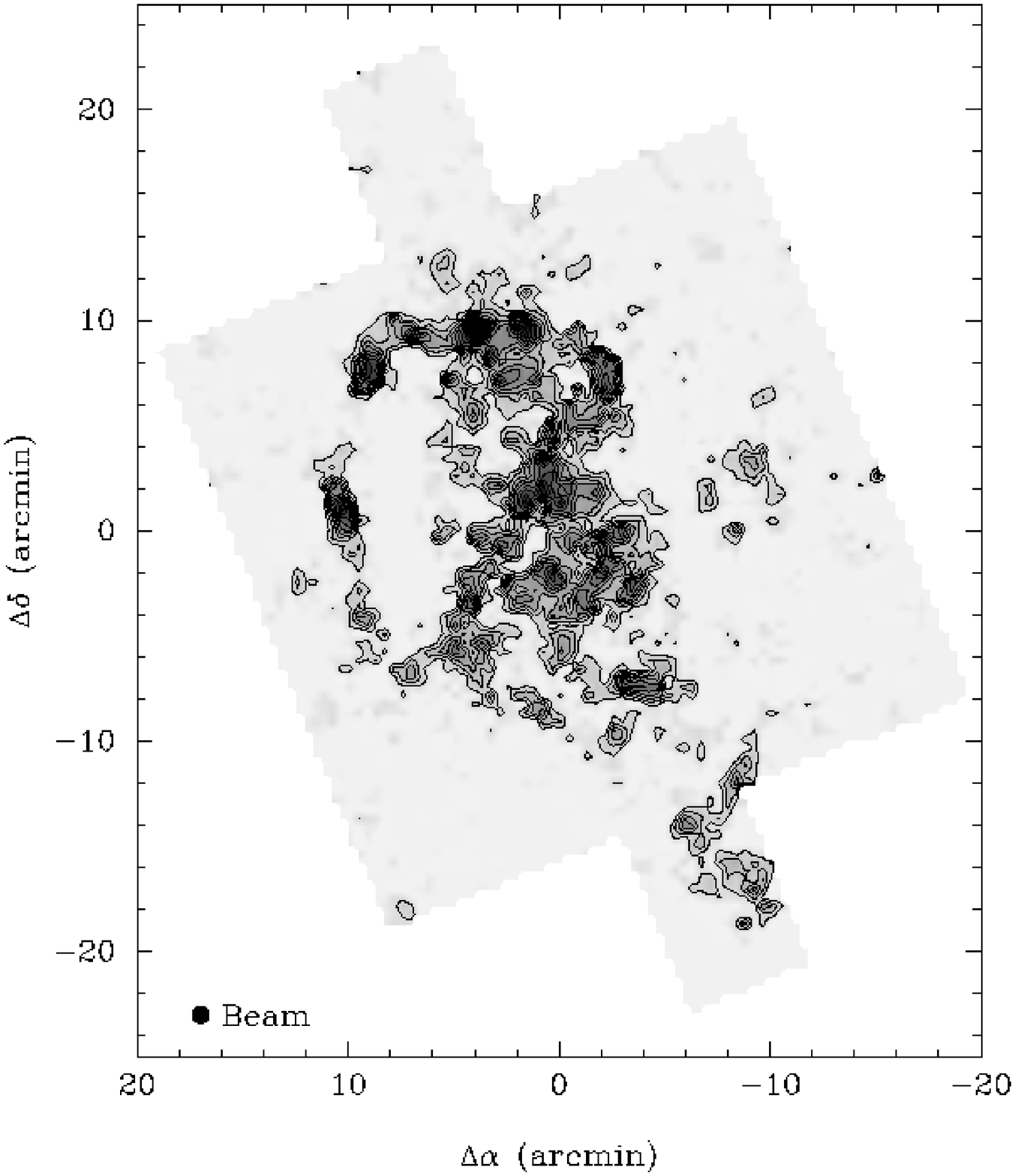}
\caption
{
A masked moment image showing the distribution of CO J=1-0 emission from M33.
The contours range from 0.6 to 4.1 K km s$^{-1}$ 
spaced by 0.5 K km s$^{-1}$
in main beam temperature units.   The shaded background shows the area
covered by these obserations. 
}
\label{co-distribution}
\end{figure}

The projected, two dimensional distribution of CO emission from M33 
as measured by these \fcrao\ observations is displayed in 
Figure~\ref{co-distribution}. The image is constructed from a masked moment 
calculation of the M33 data cube 
to minimize the noise contribution from emission free channels when integrating
over the full kinematic extent of M33 (-285 $<$ \vlsr\ $<$ -90 \kms) 
(Adler \etal 1992).  
Summing all of the emission in our map
and assuming a CO to \htwo\ conversion factor of 3$\times$10$^{20}$
cm$^{-2}$-(K kms$^{-1}$)$^{-1}$ (see Wilson \& Scoville 1990), 
we derive a global \htwo\ mass of 2.6$\times$10$^8$ \msun.  This value 
likely underestimates the molecular mass of M33 as it ignores 
the metallicity gradient within the disk (McCarthy \etal 1995; Monteverdi 
\etal 1997; Vichez \etal 1988) 
for which a larger CO to \htwo\ conversion factor may need to be applied in 
lower metallicity regions. Nor does it account for the regime 
in which \htwo\ is still abundant but CO is dissociated 
into CI or CII products.  
The bulk of CO emission is widely distributed over the 
field to a radius of 15\arcmin.
For the 50\arcsec\ resolution image, the 3$\sigma$ detection limit 
for a GMC with full 
width half maximum line width of 10 \kms\ corresponds to an \htwo\ mass of 
$\sim$8$\times 10^{4}$ M$_\odot$.
The spatial resolution of the observations (200~pc) is insufficient to detect 
individual GMCs  if their typical size is $\sim50$~pc (Wilson \& Scoville
1990; Engargiola \etal 2003; Rosolowsky \etal 2003).  Rather, the extended features 
are likely cloud structures 
distributed along spiral arms or are bound associations of GMCs.
The molecular mass is not uniformly distributed between the northern 
and southern halves.  The northern section of the disk contains 10\%
more molecular gas than the southern section.  The sense of 
this asymmetry is opposite to that measured from the HI 
distribution
over the same area
in which the ratio of atomic gas mass in 
the north is 80\% of the atomic mass in the south
 (Deul \& van der Hulst 1987). 

The degree to which giant molecular clouds reside within spiral arms provides
important clues to their origin and lifetime.
High resolution images of CO emission in more distant galaxies typically reveal
a strong concentration of molecular gas in spiral arms (Regan \etal 2001). 
In the local group, M31 exhibits a large arm to inter-arm contrast for the 
molecular gas component (Loinard \etal 1999; 
Guelin \etal 2000).  While spiral structure of molecular gas in the 
Milky Way is more difficult to determine, in the outer Galaxy where 
line of sight confusion is minimized, spiral arms are the near exclusive domain
of CO emitting clouds (Heyer \& Terebey 1998).

The spiral structure of M33 has been more difficult to define owing, 
in part, to the warp of the outer disk.  Modeling of the distribution and 
kinematics of the HI 21cm line emission demonstrates large scale warping 
for radii greater than 20\arcmin\ 
(Rogstad, Wright, \& Lockhart 1976; Corbelli \& Schneider 1997).
Similarly, Sandage \& Humphreys (1980) describe a system of spiral arms from 
the blue light distribution in which the inclination and position angles vary
from the central region to the outer disk.
Near infrared imagery shows the two bright inner spiral arms and 
identifies an enhancement of stellar density 
along the innermost spiral 
arms labeled by Sandage \& Humphreys as IS and IN
(Regan \& 
Vogel 1994).  The distribution of HI 21cm line emission from M33 
is considerably 
more complex (Newton 1980; Deul \& van der Hulst 1986).  
The optically identified arms are coincident with 
regions of enhanced atomic gas column density.  However, there are many 
radial spurs which, in projection, connect inner to outer spiral features.

\begin{figure}
\epsscale{0.9}
\plotone{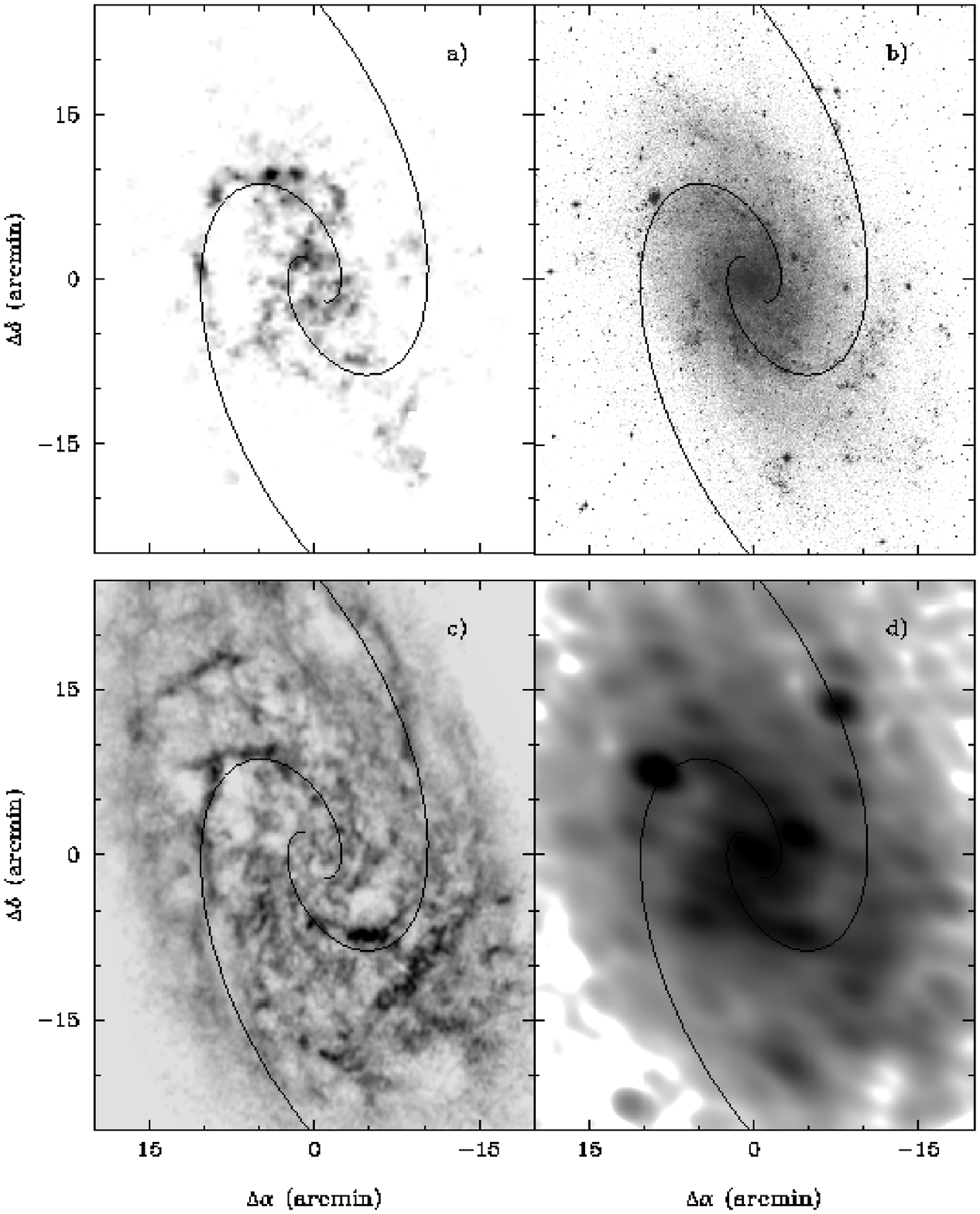}
\caption
{
Images of M33 - a) integrated $^{12}$CO J=1-0 
(halftone between 0.25 K km s$^{-1}$
(white) and 4 K km s$^{-1}$
(black), b) Digitized Sky Survey,
c) integrated HI 21cm line emission (halftone -0.1 Jy/beam m s$^{-1}$
 (white) to 0.7 Jy/beam m s$^{-1}$
(Deul \& van der Hulst 1986), and d) \IRAS\ \HIRES\ 60${\mu}m$ emission
(logarithmic stretch between 0.03 MJy sr$^{-1}$ (white) to 31.6 MJy sr$^{-1}$).
The solid line shows the spiral pattern derived
by Rogstad, Wright, \& Lockhart (1976).  
Most of the CO emission can be assigned to the
prominent north and south spiral arms.
}
\label{co_dss_hi}
\end{figure}

Figure~\ref{co_dss_hi} shows the the distribution of CO emission, the 
Digitized Sky Survey image, 
HI 21cm line emission from Westerbork Radio Synthesis Telescope
(Deul \& van der Hulst 1986), and \IRAS\ \HIRES\ 60$\mu$m emission.  
The spiral pattern derived by Rogstad, Wright,
\& Lockhart (1976) is overlayed as a guide to associate the 
gas distribution with some plausible spiral structure in M33.
While this two arm pattern with a wide opening 
angle provides a reasonable fit to the brightest HI and CO spiral arm
segments,  it is not a unique solution nor 
does it accurately describe all of the neutral 
gas emission.  In particular, the  HI and CO segments at
offsets (-7\arcmin,-13\arcmin), (+4\arcmin,-7\arcmin), and the 
isolated, northern HI segment at (+10\arcmin,+16\arcmin) are well displaced
from the model spiral arms.  These are likely spiral features that do not
conform to a global spiral pattern.

\subsection{Molecular Gas Fraction}
The gas component of disk galaxies is cycled between molecular, atomic, 
and ionized phases of the ISM.  Molecular clouds condense from the reservoir 
of atomic gas in regions with sufficient column density to self-shield molecules
from dissociating UV radiation.
In regions of star formation where the UV radiation field is high, 
molecular material can be photo-dissociated or photo-ionized.  The 
fraction of molecular gas relative to the atomic gas component and the 
radial profile of this ratio provides insight to the recycling process.

\begin{figure}
\plotone{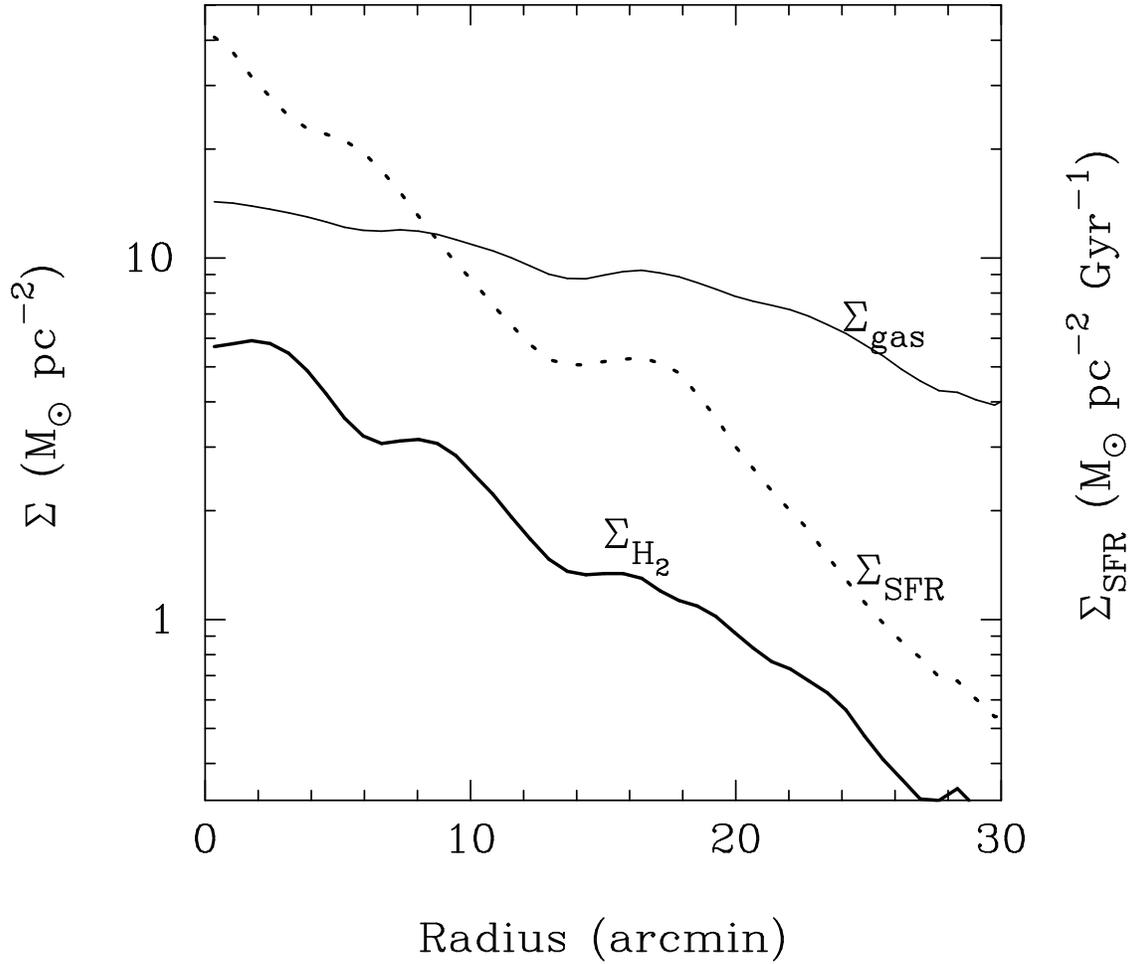}
\caption { The radial profile of molecular gas surface density, $\Sigma_{H_2}$ (heavy solid 
line), the total gas surface density, $\Sigma_{gas}$ (light solid line), and the 
star formation rate per unit area, $\Sigma_{SFR}$ (dotted line).}
\label{radial}
\end{figure}

\begin{figure}
\plotone{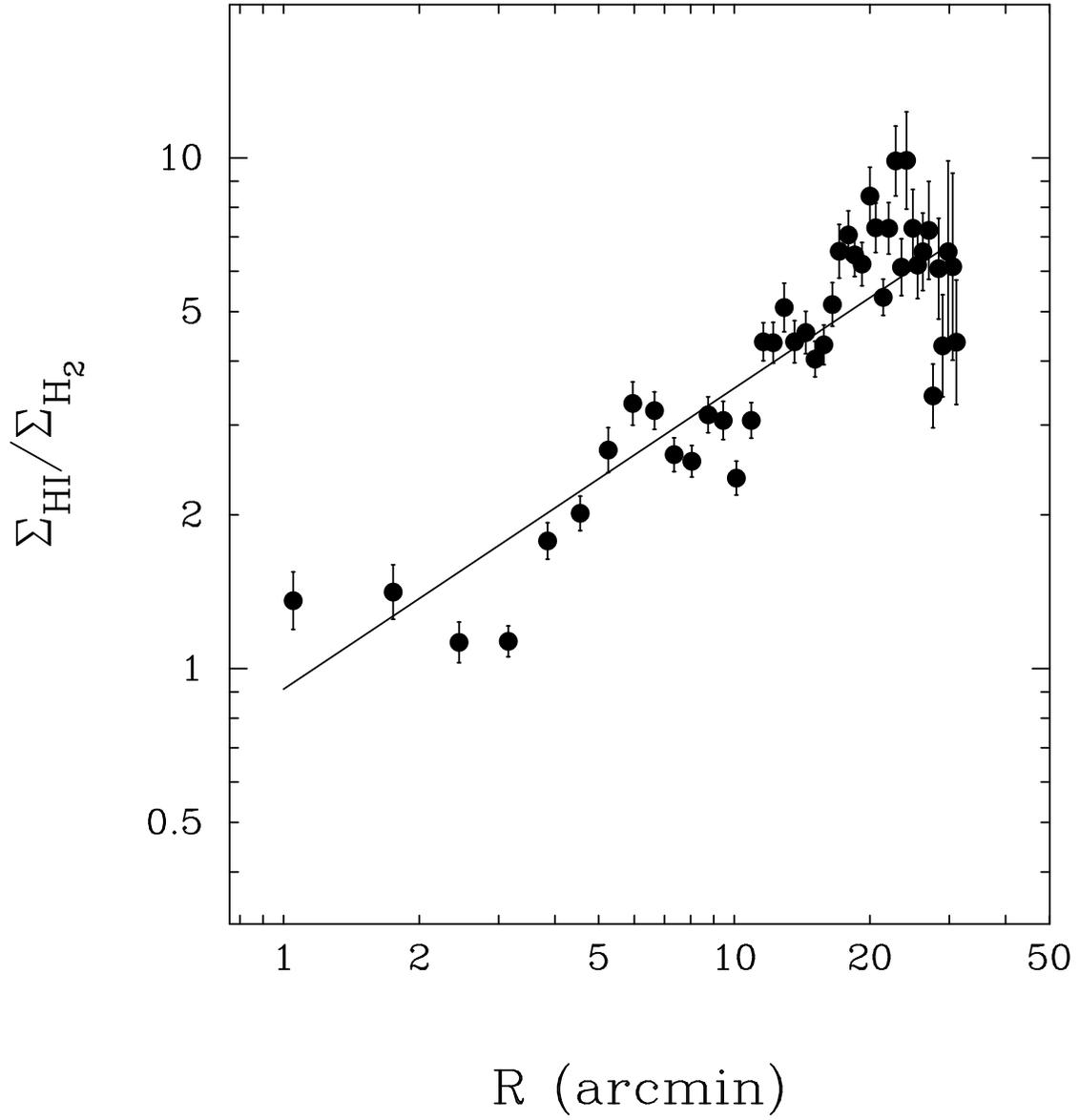}
\caption { The radial profile of the ratio of 
atomic to molecular hydrogen mass surface densities.  The solid line 
is the best fitting power law, $\Sigma_{HI}/\Sigma_{H_2}=0.9\times R^{0.6}$. }
\label{hi_h2}
\end{figure}

Corbelli (2003) has derived the radial profiles of the molecular and 
atomic gas surface densities in M33 from the \fcrao\ CO data and the Arecibo  
21-cm data (FWHM resolution 3.9\arcmin) using a tilted ring model which best fit the atomic hydrogen
distribution (Corbelli \& Schneider 1997).  The same tilted ring model is 
applied to 
the 21-cm Westerbork image of M33  
(Deul \& van der Hulst 1987) smoothed to the same spatial resolution (50\arcsec)
to derive 
the radial
dependence of the ratio of the atomic to molecular gas mass surface density,
$\Sigma_{HI}/\Sigma_{H_2}$.  The Westerbork interferometer image recovers 
64\% of the HI flux within a radius of 20\arcmin\ and 60\% of the flux
within a radius of 30\arcmin.
Figure~\ref{radial} shows the radial profiles of $\Sigma_{H_2}$ and 
$\Sigma_{gas}$ 
using the 
Westerbork data with no correction for missing flux.
Figure~\ref{hi_h2} shows the variation of $\Sigma_{HI}/\Sigma_{H_2}$ with radius.
Errors on $\Sigma_{HI}$ and $\Sigma_{H_2}$ are 
computed according to the dispersion around the mean in each ring. 
The atomic gas component dominates the total gas surface density at all 
radii greater than 4\arcmin.  
The radial dependence for the HI/H$_2$ ratio is reasonably described by 
the power law,
$$\Sigma_{HI}/\Sigma_{H_2}\propto R^{0.6} \eqno (3) $$ 
Annular averages derived using the Arecibo 21-cm data (Corbelli \& Schneider
1997) provide a similar best fitting power law index and suggests that 
the fraction of flux recovered by the Westerbork interferometer does not 
vary with radius.
The radial slope of $\Sigma_{HI}/\Sigma_{H_2}$ is  
shallower in M33 than in gas rich galaxies for which the 
power law index is $\sim$1.5 (Wong \& Blitz 2002).

A point to point comparison of the CO and HI images show varying degrees
of correlation. 
In the central 4\arcmin\ of M33, there is bright CO emission but 
only 
low to moderate HI surface brightness indicating an efficient conversion
from atomic to molecular gas.  Outside of this central region,
most of the detected CO emission is coincident with high brightness
HI features.
From a sample of high resolution VLA images of 11 nearby spiral galaxies,
Braun (1997) describes the HI distributions as a network of high brightness
emission and associates this material with the 
cold, neutral component of the ISM based on the observed 
narrow HI line widths.  The correspondence between CO emission 
and high brightness HI filaments in the disk of M33 demonstrates the 
condensation of giant molecular clouds from the cold, atomic gas but with 
lower efficiency than is observed in the central 4\arcmin.

\section{Discussion}

\subsection{The Schmidt Law in M33}
The survey of CO emission from M33 presented in \S3 provides the first 
unbiased, filled aperture census of molecular gas in this Local 
Group galaxy. Such accounting of the molecular gas phase is essential 
to construct accurate descriptions of the star formation process 
within the disk.  
The Schmidt Law provides a concise, albeit, oversimplified recipe for 
star formation within the disk of a galaxy.  It is typically derived from 
the radial profiles of the gas surface density (atomic + molecular)
and some measure of the star formation rate, typically, H$\alpha$ or 
\IRAS\ far infrared flux densities (Kennicutt 1989).  
From a sample of 15 galaxies, 
Kennicutt (1989) determined the mean power law index, $n$, to be 1.3$\pm$0.3.
Recent high resolution observations of galaxies derive a Schmidt Law index 
between 1.1 and 1.7 depending on the extinction correction
 applied to the H$\alpha$
(Wong \& Blitz 2002).  More importantly, Wong \& Blitz (2002) note that the 
star formation rate surface density is more strongly correlated with the 
\htwo\ surface density than the total gas surface density as one would expect
given that stars condense from giant molecular clouds.  This confirms 
previous studies of the face-on spiral galaxies M51 and NGC~6946 where 
the H$\alpha$ emission is strongly correlated with the distribution of 
molecular gas (Lord \& Young 1990; Tacconi \& Young 1990).  However, at the 
highest resolution, the brightest HII regions are not coincident with 
CO emitting regions as the molecular gas is either photo-dissociated
or ionized by the strong UV radiation from massive stars or evacuated by 
the attendant stellar winds. 

Previous efforts to derive the Schmidt Law in M33 have considered the 
atomic gas exclusively or have been restricted to the 
central 3.5\arcmin\ owing to the limited imaging of the molecular 
gas component.  Madore \etal (1974) and Newton (1980) derived power law 
relationships between the surface density of HII regions and the atomic 
gas column density from interferometric HI 21cm line images.
Newton (1980) found a nearly flat relationship between the two quantities in
the inner regions of M33 ($n\sim$0.6) and a steep power law ($n\sim$2)
in the outer regions (R $>$ 2.4 kpc).  

To gauge the star formation rate in M33, we use the far infrared luminosity, 
$L_{FIR}$, 
derived from \IRAS\ \HIRES\ 60\micron\ and 100\micron\ images of M33.
The far infrared emission provides an excellent tracer to 
star formation activity in galaxies as a substantial amount of the 
far infrared flux emitted by a galaxy 
is contributed by interstellar dust heated by massive,
newborn stars.  Moreover, the far infrared emission is optically thin
throughout the disk making corrections for extinction 
unnecessary.  The old stellar component of a galaxy may also 
contribute to the measured far infrared flux.  For M33, this contribution 
has been estimated to be less than 30\% (Devereux, Duric, \& Scowen 1997).

Owing to the deconvolution process, the resolution of the \HIRES\ 
images may vary across the field.  For the central 30\arcmin\ of the M33 \HIRES\
field, the resolutions of the 
60\micron\ and 100\micron\ images are directly 
 estimated from the ancillary beam maps and found to vary between 4 and 10\%.
To assure uniform, circular point spread functions of the 60\micron\ and 
100\micron\ images with which to compare the CO and WRST HI images, 
the far infrared data were smoothed to a final 
resolution of 120\arcsec\ using gaussian kernels with full width 
half maxima of 98\arcsec\ and 48\arcsec\ respectively.  The \fcrao\ CO and 
Westerbork  HI 
images were similarly smoothed to this resolution. 

The star formation rate, $dM_*/dt$,
 within a given solid angle and averaged over a
characteristic time is derived from the measured far infrared luminosity,
$$  { {dM_*} \over {dt} } = 
6.3{\times}10^{-10} L_{FIR} \;\;\;\; M_\odot yr^{-1} \eqno(3) $$
where
$$  L_{FIR} = 6{\times}10^5 D^2(2.58 \int I_{60} d\Omega + 
\int I_{100}d\Omega) \;\; L_\odot,  \;\;\;\; \eqno(4) $$
D is the distance to M33 (0.83 Mpc), and a Miller-Scalo IMF is assumed
 (Thronson \& Telesco 1986).
The star formation rate per unit area, $\Sigma_{SFR}$, is 
$$ \Sigma_{SFR}={ {3.8{\times}10^{-4}D^2(2.58 \int I_{60} d\Omega + 
\int I_{100}d\Omega)} \over {10^{12} D^2 \int d\Omega} }  \;\;\; 
M_\odot pc^{-2} yr^{-1} \;\;\; \eqno(5) $$
$$ \Sigma_{SFR}=3.8{\times}10^{-16}(2.58 <I_{60}>_\Omega+<I_{100}>_\Omega) 
\;\;\; M_\odot pc^{-2} yr^{-1} \eqno(6) $$
where $<I_{60}>_\Omega$ and $<I_{100}>_\Omega$ are the mean 60\micron\ 
and 100\micron\ intensities within a solid angle, $\Omega$, respectively.  
Many studies  use the annular method in which 
$\Sigma_{SFR}$ is tabulated within the solid angle subtended by 
the annulus of a 
de-projected tilted ring and the gas surface density is the mean surface
density within this ring (Kennicutt 1989; Martin \& Kennicutt 2001; 
Wong \& Blitz 2002).  A Schmidt Law is constructed by evaluating 
$\Sigma_{SFR}$ and $\Sigma_{gas}$ within successive rings at varying radii.
The annular method provides a time averaged estimate of $\Sigma_{gas}$ 
as it does not consider local variations of the gas distribution and 
phase caused by the star formation process itself.  
Following Wong \& 
Blitz (2002), errors to $\Sigma_{SFR}$ and $\Sigma_{gas}$ are assigned to 
the standard deviation of values within the subtended area.  
The radial
profile of $\Sigma_{SFR}$
is shown in Figure~\ref{radial}.

\begin{figure}
\plotone{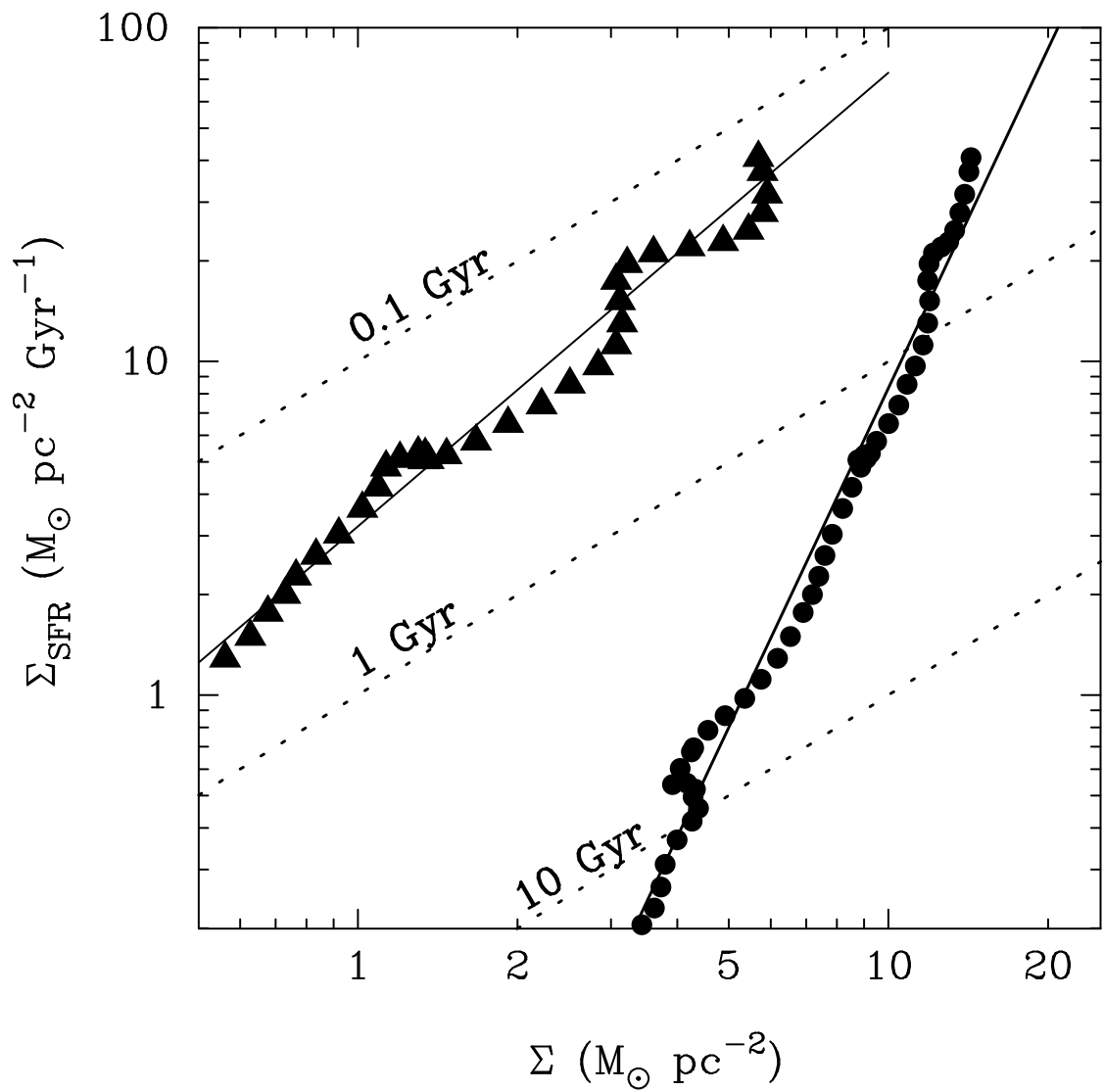}
\caption {The variation of the star formation rate per unit area with 
total (circles) and molecular (triangles) gas surface densities.
The solid lines show the power law fits to the data.
The dashed lines show the gas depletion time scales.}
\label{schmidt}
\end{figure}

The variation of star formation rate per unit 
area with both molecular and total gas surface densities in M33 
is shown in Figure~\ref{schmidt}. 
Power laws are fit to each type of 
surface density and the amplitudes and power law indices are listed in Table~1.
The fits are weighted by errors in both 
$\Sigma_{SFR}$ and $\Sigma_{gas}$. 
The steep relationship found for the total gas surface density 
is due to the near constant profile of the atomic gas 
component in the central 12\arcmin\ which dominates the total gas surface 
density for all radii greater than 4 \arcmin\ 
(Corbelli 2003).   

There is a strong correlation between the derived star formation rate per unit area
and molecular gas surface density.  
This relationship even extends into the regime 
in which the neutral gas is mostly atomic.
Such a correlation is not an artifact due to 
local heating of the molecular gas from an enhanced UV radiation field.
An examination of the far infrared and CO emission from 
molecular clouds in the solar neighborhood show that most 
of the far infrared luminosity emerges from localized star forming cores that cover a small
fraction of the cloud area.
Beyond several pc, the UV
radiation from young, massive stars provides little or no
heating to the extended gas distribution from which most of the CO luminosity originates.
Rather the measured correlation between $\Sigma_{SFR}$ and $\Sigma_{H_2}$ is a testament 
that stars exclusively condense from clouds of molecular gas.
This relationship 
 suggests that the formation of 
molecular clouds is the essential prerequisite to star formation (see Engargiola \etal
2003).
The molecular Schmidt index, $n_{mol}=1.36\pm0.08$, 
is similar to values determined by
Wong \& Blitz (2002) when radially dependent extinction corrections are applied
to their H$\alpha$ data.  
The star formation rates at all radii are sufficiently high that if 
sustained, the current reservoir of molecular gas would be depleted within 
1-3$\times$10$^8$ years.  The depletion time for the total gas content increases 
from 0.4 in the central regions to greater than 10 Gyr in the outer disk where 
the star formation rate is reduced. Such variations are commonly 
observed in other galaxies (Wong \& Blitz 2002).  To sustain such star formation rates in the inner disk 
of M33 beyond several Gyr
requires funneling gas from the outer disk reservoir or a shallow
initial mass function in which a larger fraction of massive stars are formed
to replenish the interstellar medium.   

\subsection{The Transition from Atomic to Molecular Gas}

The major difference between M33 and more luminous 
gas rich galaxies is the relationship between atomic and
molecular gas. Galaxies in the sample of Wong \& Blitz (2002)
are molecular-rich such that the atomic to molecular gas ratio, 
$\Sigma_{HI}/\Sigma_{H_2}$, ranges between 0.05 and 0.4 
and varies with galactocentric radius 
roughly as $R^{1.5}$.  In M33,  
$\Sigma_{HI}/\Sigma_{H_2}$ varies between 1 and 10 with a radial 
dependence,  $R^{0.6}$ (see Figure~\ref{hi_h2}). 

In order to explain this radial dependence, we
consider the model proposed by Elmegreen (1993).
This model is based on the assumption that the fraction of molecular
gas is determined by the balance between the gas pressure,
which regulates the $H_2$ formation rate, and the dissociating 
radiation. In particular
the molecular gas fraction, $f_{mol}$, should depend upon the 
the ambient pressure, $P$, and radiation field, $j$, as 
$$f_{mol} \equiv {\Sigma_{H_2}\over \Sigma_{gas}} \propto 
P^{2.2} j^{-1}$$  
To evaluate this interplay between the pressure and radiation field, 
the pressure 
of a disk in hydrostatic equilibrium is approximated as
$$P\simeq {\pi \over 2}G\Sigma_{gas}\Bigl(\Sigma_{gas}+{c_{gas}\over c_{star}}
\Sigma_{star}\Bigr) \eqno(7) $$
$\Sigma_{star}$ is the
star surface density of the disk which best fits the rotation curve
(Corbelli 2003, Figure 5$(b)$). The velocity dispersion of the gaseous and
stellar disk are $c_{gas}=8$~km~s$^{-1}$ and 
$c_{star}=(\pi G z_0 \Sigma_{star})^{1/2}$, consistent with an isothermal disk 
of scale height $z_0=0.5$~kpc.
Using $j\propto \Sigma_{star}$,  we derive $\alpha$ that best fits the
relation,
$${j\over j_0} f_{mol}= f_0 \Bigl({P\over P_0}\Bigr)^{\alpha} \eqno(8) $$
where $f_0$, $j_0$ and $P_0$ are the values of $f_{mol}$, $j$ and $P$ in
the innermost ring. By minimizing the $\chi^2$ value we find, 
$\alpha=2.3\pm0.05$ where 
the quoted uncertainty is the 1$\sigma$ error on the best fitting value of
$\alpha$ that takes into account $f_{mol}$ uncertainties only.
The Pearson linear correlation coefficient is 0.97.
The data in M33 are in good agreement with Elmegreen's predictions
and parameterization for galaxies with low molecular gas fractions.
These empirically support the idea that pressure is playing 
a significant role in regulating the $H_2$ fraction in low luminosity galaxies, 
such as M33. 

While M33 is quite distinct from the sample of galaxies studied
by  Wong \& Blitz (2002) 
in terms of gas surface density, Q parameters, and molecular 
gas fraction, it adheres to a similar  star formation recipe.  That is, 
star formation is primarily limited by the condensation of giant molecular 
clouds from the diffuse interstellar medium.  Once a GMC develops, star 
formation is rapidly initiated (Engargiola \etal 2003). 
If $H_2$ regulates the star formation rate, the high 
Schmidt index, $n$, relative to the total gas surface density, can be explained 
as a consequence of $n_{mol}$ and of the molecular gas fraction, $f_{mol}$. 
Following Wong \& Blitz (2002), the Schmidt 
index, $n$, can be written as
$$n=n_{mol}\Bigl({d {\hbox{ln}} \Sigma_{H_2} \over d {\hbox{ln}} 
\Sigma_{gas}}\Bigr) \eqno(9) $$
From the annular averages we find $d{\hbox{ln}} \Sigma_{H_2} / d{\hbox{ln}} 
\Sigma_{gas} = 2.3\pm 0.05$. Since $n_{mol}=1.36\pm 0.08$, 
the above equations predicts $n=3.1\pm$0.2.  Within the errors, this two step
model of star formation is consistent with the 
observed value $n=3.3\pm0.1$.
The atomic gas layer of the disk may ultimately
provide the reservoir of material from which giant molecular clouds
condense and form stars in M33 but it plays no obvious role in regulating 
star formation.

\subsection{Radial Progression of Star Formation in M33}
Comparisons of the molecular scale length with the disk scale
lengths derived from the stellar light,
HII regions, and OB associations  
provides coarse insight to the star formation history of a galaxy.
Corbelli (2003) derived a molecular gas scale 
length of $\sim$11\arcmin\ (2.6 kpc).
The disk scale lengths of OB associations and HII regions are comparable
but just interior to this CO scale length and attest to the strong coupling 
of the molecular gas and recent ($<$10$^7$ years) star formation activity.  
Star formation in the more distant past is reflected in the distribution 
of the stellar light within the disk of a galaxy.
From J, H, K band imaging, Regan \& Vogel (1994) derive a disk 
stellar scale length of order 6\arcmin.   Thus, the disk stellar light, 
particularly, the older population of stars, is much more centrally condensed
than the molecular gas.   
The broader distribution of CO in M33 with respect to old populations of stars 
is not typical of spiral galaxies.  From the BIMA SONG,
4 of the 15 galaxies with measured profiles show CO scale lengths significantly 
(3$\sigma$) 
larger than the stellar scale length and one galaxy (NGC~628) with 
a difference as large as M33 (Regan \etal 2001).
van den Bergh (2000) noted that the broader distribution of HII regions
with respect to the starlight reflects an outward progression of 
star formation with time.  As the molecular gas distribution 
provides a roadmap for both current and future star formation within a
galaxy, the large molecular scale length with respect to the stellar 
light provides additional evidence for
the outward progression of star formation activity in M33.

\section{Conclusion}

In this paper we have presented an image of  $^{12}$CO J=1-0 emission from M33
and investigated the relationship between the star formation 
rate and the molecular and atomic neutral gas components.  
From our analyses, we derive the following conclusions.
\begin{enumerate}
\item The mass surface density of molecular gas is strongly correlated 
with the star formation rate per unit area even in regions that 
are dominated by atomic gas.  This proportionality suggests that the 
formation of molecular clouds is the limiting step to the formation of 
stars within the disk of M33.  
\item The Schmidt indices for the total and molecular gas surface densities are
$n=3.3\pm0.1$ and $n_{mol}=1.36\pm0.08$ respectively.  The large value for $n$ is a
consequence of the variation of the star formation rate within the molecular gas 
substrate, ($n_{mol}$), and the condensation of molecular clouds from the 
atomic gas component
($f_{mol}$).
\item 
The pressure of the gaseous disk layer due to the weight of both 
gas and stars can account for the 
radial variations of the molecular gas fraction.
\item  The distributions of molecular gas and current star formation activity are 
extended with respect to the old, stellar population and suggests an outward 
progression of star formation within the M33 disk.

\end{enumerate}

\acknowledgments
We thank Thijs van der Hulst for providing us 
with the Westerbork HI 21cm line image.
Insightful comments from Erik Rosolowsky and 
the referee are appreciated.
This work was supported by NSF grant AST 02-28993  to the Five College
Radio Astronomy Observatory.

\clearpage

\clearpage
\begin{table}[htb]
\begin{center}
\caption{Observed Schmidt Law $\Sigma_{SFR}=C\Sigma_{gas}^n$ in M33}
\vspace{7mm}
 \begin{tabular}
{lcc}
\hline
& Total Gas & Molecular Gas  \\
 \hline
n & 3.3$\pm$0.07& 1.36$\pm$0.08\\
C & 0.0035$\pm$ 0.066 & 3.2$\pm$0.2 \\
Corr. Coeff. & 0.99 & 0.98 \\
\hline
 \end{tabular}
 \end{center}
\label{indices}
\end{table}
\end{document}